\documentclass[aps,a4paper,superscriptaddress,preprintnumbers,showpacs,
amsmath,amssymb]{revtex4}
\usepackage{graphicx}
\usepackage{dcolumn}
\usepackage{color}
\usepackage{latexsym,amsfonts}
\usepackage{textcomp}
\usepackage{bm}

\usepackage{ulem}

\begin{document}

\title{Can chameleon field be identified with quintessence ?}

\author{A. N. Ivanov}\email{ivanov@kph.tuwien.ac.at}
\affiliation{Atominstitut, Technische Universit\"at Wien, Stadionallee
  2, A-1020 Wien, Austria}
\author{M. Wellenzohn}\email{max.wellenzohn@gmail.com}
\affiliation{Atominstitut, Technische Universit\"at Wien, Stadionallee
  2, A-1020 Wien, Austria} \affiliation{FH Campus Wien, University of
  Applied Sciences, Favoritenstra\ss e 226, 1100 Wien, Austria}

\date{\today}

\begin{abstract}
In the Einstein--Cartan gravitational theory with the chameleon field,
changing its mass in dependence of a density of its environment, we
analyze the Friedmann--Einstein equations for the Universe evolution
with the expansion parameter $a$ dependent on time only. We analyze
the problem of an identification of the chameleon field with
quintessence, i.e. a canonical scalar field responsible for dark
energy dynamics and for the acceleration of the Universe expansion. We
show that since the cosmological constant, related to the relic dark
energy density, is induced by torsion (Astrophys. J. {\bf 829}, 47
(2016)), the chameleon field may, in principle, possess some
properties of quintessence such as an influence on the dark energy
dynamics and the acceleration of the Universe expansion even on the
late-time acceleration, but it cannot be identified with quintessence
to full extent.
\end{abstract}

\maketitle

\section{Introduction}
\label{sec:introduction}

The chameleon field, changing its mass in dependence of a density of
its environment \cite{Khoury2004,Mota2007}, has been invented to avoid
the problem of the equivalence principle violation
\cite{Will1993}. Nowadays it is accepted that the chameleon field,
identified with quintessence \cite{Ratra1988} -- \cite{Tsujikawa2013},
i.e. a canonical scalar field, can be useful for an explanation of the
late--time acceleration of the Universe expansion
\cite{Perlmutter1997,Riess1998,Perlmutter1999,Goobar2000} and may shed
light on the origin of dark energy and a dark energy dynamics
\cite{Peebles2003}--\cite{Pignol2015}. Since the relic dark energy
density is closely related to the cosmological constant
\cite{Peebles2003}, in contrast to such a hypothesis that the
chameleon field might originate the cosmological constant,
proportional to the homogeneous static dark energy density, there has
been shown at the model--independent level within the Einstein--Cartan
gravitational theory \cite{Cartan1922} -- \cite{Ni2010} that the
cosmological constant or the relic dark energy density has a
geometrical origin caused by torsion \cite{Ivanov2016a}. In this case
the chameleon field is able only to evolve above the relic background
of the dark energy simulating its dynamics and, of course, to make a
certain influence on the acceleration of the Universe expansion.

For the observation of torsion in the terrestrial laboratories there
have been derived potentials of low--energy torsion-neutron
interactions \cite{Ivanov2015a} -- \cite{Ivanov2016}. In terrestrial
laboratories extreme smallness of absolute values of torsion was
confirmed in different estimates of constraints on contributions of
torsion to observables of elementary particle interactions
\cite{Laemmerzahl1997} -- \cite{Ivanov2016d}, including the qBounce
experiments with ultracold neutrons \cite{Abele2010} --
\cite{Jenke2014} (see also \cite{Ivanov2016d}).

The chameleon--matter interactions were also intensively investigated
in terrestrial laboratories \cite{Abele2010}--\cite{Li2016} in
experiments with ultracold and cold neutrons through some effective
low--energy chameleon--neutron potentials
\cite{Brax2011}--\cite{Ivanov2014} and by using cold atoms in the atom
interferometry \cite{Burrage2014, Burrage2015, Hamilton2015,
  Hamilton2016, Burrage2018}. However, recently an importance of the
chameleon field as quintessence in the late--time acceleration of the
Universe has been questioned by Wang {\it et al.}  \cite{Wang2012} and
Khoury \cite{Khoury2013} by pointing out that the conformal factor,
relating the Einstein and Jordan frames and defining the
chameleon--matter interactions, is essentially constant over the last
Hubble time. According to Wang {\it et al.}  \cite{Wang2012} and
Khoury \cite{Khoury2013}, this implies a negligible influence of the
chameleon field on the late--time acceleration of the Universe
expansion. To some extent this should also imply that the chameleon
field cannot possess such a property of quintessence as a
responsibility for the late--time acceleration of the Universe
expansion \cite{Peebles1999a} -- \cite{Steinhardt2001a}.

Thus, the aim of this paper is to investigate the properties of the
chameleon field in comparison to the properties of quintessence. Below
we show that the chameleon field has no relation to the origin of the
cosmological constant or the relic dark energy density, which is
induced by torsion \cite{Ivanov2016a}. However, the chameleon field
can still influence on the Universe expansion even on the late--time
acceleration, caused by its evolution above the background of the
relic dark energy \cite{Ivanov2016a}. By analysing the Einstein
equations for the flat Universe in the spacetime with the Friedmann
metric, dependent on the expansion parameter $a$ \cite{Rebhan2012}, we
show that conservation of a total energy--momentum tensor of the
system, including the chameleon field, radiation and matter (dark and
baryon matter), demands the conformal factor to be equal to unity if
and only if the dependence of the radiation $\rho_r(a)$ and matter
$\rho_m(a)$ densities on the expansion parameter $a$ does not deviate
from their standard form $\rho_r(a) \sim a^{-4}$ and $\rho_m(a) \sim
a^{-3}$, respectively \cite{Rebhan2012}.  The same result we obtain by
analysing the first order differential Friedmann--Einstein equation,
relating $\dot{a}^2/a^2$ to the chameleon field, radiation and matter
densities, and the second order differential Friedmann--Einstein
equation, relating $\ddot{a}/a$ to the chameleon field, radiation and
matter densities and their pressures, where $\dot{a}$ and $\ddot{a}$
are the first and second time derivatives of the expansion parameter.
Of course, the equality of the conformal factor to unity suppresses
any coupling of the chameleon field to a matter density of its
environment and makes such a scalar field unuseful to avoid the
problem of the equivalence principle violation
\cite{Will1993}. However, it does not prevent the chameleon field,
evolving above the background of the relic dark energy, from a
simulation of a dark energy dynamics and an influence on an
acceleration of the Universe expansion. Then, we show that the
Friedmann--Einstein equation for $\dot{a}^2/a^2$ is the first integral
of the Friedmann--Einstein equation for $\ddot{a}/a$ if and only if
the total energy--momentum of the system, including the chameleon
field, radiation and matter, is locally conserved. As a result we
infer that i) if the radiation and matter densities obey their
standard dependence on the expansion parameter $\rho_r(a) \sim a^{-4}$
and $\rho_m(a) \sim a^{-3}$ the conformal factor is equal to unity and
the chameleon field loses a possibility to couple to an environment,
and ii) if the dependence of the radiation and matter densities
deviate from their standard behaviour $\rho_r(a) \sim a^{-4}$ and
$\rho_m(a) \sim a^{-3}$ the conformal factor is not equal to unity and
makes possible interactions of the chameleon field with its
environment. In this case an usage of the chameleon field to the
problem of the equivalences principle violation becomes meaningful. In
spite of the fact that the chameleon field does not possess the main
property of quintessence to be a hypothetical form of dark energy
\cite{Ratra1988}, since the relic dark energy density or the
cosmological constant has a geometrical origin related to torsion
\cite{Ivanov2016a}, the chameleon field, evolving above the relic dark
energy and simulating a dark energy dynamics, might be responsible for
an acceleration of the Universe expansion.

The paper is organized as follows. In section \ref{sec:ecaction} we
derive the Einstein equations in the Einstein--Cartan gravitational
theory with torsion, chameleon and matter fields. Following
\cite{Ivanov2016a} we show that the contribution of torsion to the
Einstein--Hilbert action is presented in the form of the cosmological
constant. Then, following Khoury and Weltman \cite{Khoury2004} we
include the part of the integrand of the Einstein--Hilbert action
proportional to the cosmological constant to the potential of the
self--interaction of the chameleon field. This implies that the
chameleon field has no relation to an origin of the cosmological
constant or the relic dark energy density but can only evolve above
such a relic background caused by torsion and simulate a dark energy
dynamics. In section \ref{sec:evolution} in the flat Friedmann
spacetime with the standard Friedmann metric $g_{\mu\nu}$,
i.e. $g_{00} = 1$, $g_{0j} = 0$ and $g_{ij} = a^2(t)\,\eta_{ij}$ and
$\eta_{ij} = - \delta_{ij}$, we show that the Einstein equations
reduce themselves to the Friedmann--Einstein equations of the Universe
evolution with the chameleon field, radiation and matter (dark and
baryon) densities. Since the Einstein tensor $G_{\mu\nu} = R_{\mu\nu}
- \frac{1}{2}\,g_{\mu\nu}R$, where $R_{\mu\nu}$ and $R$ are the Ricci
tensor and scalar curvature, respectively, obey the Bianchi identity
$G^{\mu\nu}{}_{;\mu} = 0$, where $G^{\mu\nu}{}_{;\mu}$ is the
covariant divergence \cite{Rebhan2012}, the total energy--momentum
tensor of the system, including the chameleon field, radiation and
matter (dark and baryon) should be locally conserved. We find that
local conservation of the total energy--momentum tensor imposes the
evolution equations for the radiation and matter densities, where the
dependence of which on the expansion parameter $a$ is corrected by the
conformal factor in comparison to the standard dependence $\rho_r(a)
\sim a^{-4}$ and $\rho_a \sim a^{-3}$, respectively
\cite{Rebhan2012}. We show that the Friedmann--Einstein equation for
$\dot{a}^2/a^2$ is the first integral of the Friedmann--Einstein
equation for $\ddot{a}/a$ if and only if the total energy momentum of
the system, including the chameleon field, radiation and matter, is
locally conserved. In case of the standard dependence of the radiation
and matter densities on the expansion parameter $\rho_r(a) \sim
a^{-4}$ and $\rho_m \sim a^{-3}$ \cite{Rebhan2012} local conservation
of the total energy--momentum tensor of the chameleon field, radiation
and matter demands the conformal factor to be equal to unity. This
suppresses any interaction of the chameleon field with an ambient
environment. In section \ref{sec:Schluss} we discuss the obtained
results.

\section{Einstein's equations in the Einstein--Cartan gravitational
  theory with chameleon and matter fields}
\label{sec:ecaction}

The Einstein--Hilbert action of the Einstein--Cartan gravitational
theory without chameleon and matter fields we take in the standard
form \cite{Schroedinger1950, Kostelecky2004, Rebhan2012}
\begin{eqnarray}\label{eq:1}
S_{\rm EH} = \frac{1}{2}\,M^2_{\rm Pl}\,\int d^4x\,\sqrt{-g}\,{\cal R},
\end{eqnarray}
where $M_{\rm Pl} = 1/\sqrt{8\pi G_N} = 2.435\times 10^{27}\,{\rm eV}$
is the reduced Planck mass, $G_N$ is the Newtonian gravitational
constant \cite{PDG2020} and $g$ is the determinant of the metric
tensor $g_{\mu\nu}$. The scalar curvature ${\cal R}$ is defined by
\cite{Kostelecky2004, Schroedinger1950}
\begin{eqnarray}\label{eq:2}
{\cal R} = g^{\mu\nu} {\cal R}^{\alpha}{}_{\mu\alpha\nu} = g^{\mu\nu}
\Big(\frac{\partial}{\partial x^{\nu}}\Gamma^{\alpha}{}_{\alpha\mu} -
\frac{\partial}{\partial x^{\alpha}}\Gamma^{\alpha}{}_{\nu\mu} +
\Gamma^{\alpha}{}_{\nu\varphi} \Gamma^{\varphi}{}_{\alpha\mu} -
\Gamma^{\alpha{}}_{\alpha\varphi}\Gamma^{\varphi}{}_{\nu\mu}\Big) =
g^{\mu\nu} {\cal R}_{\mu\nu},
\end{eqnarray}
where ${\cal R}^{\alpha}{}_{\mu\beta\nu}$ and ${\cal R}_{\mu\nu}$ are
the Riemann and Ricci tensors in the Einstein--Cartan gravitational
theory, respectively. Then, $\Gamma^{\alpha}{}_{\mu\nu}$ is the affine
connection
\begin{eqnarray}\label{eq:3}
\Gamma^{\alpha}{}_{\mu\nu} = \{^{\alpha}{}_{\mu\nu}\} + {\cal
  K}^{\alpha}{}_{\mu\nu} = \{^{\alpha}{}_{\mu\nu}\} +
g^{\alpha\sigma}{\cal K}_{\sigma\mu\nu},
\end{eqnarray}
where $\{^{\alpha}{}_{\mu\nu}\}$ are the Christoffel symbols
\cite{Rebhan2012}
\begin{eqnarray}\label{eq:4}
\{^{\alpha}{}_{\mu\nu}\} =
\frac{1}{2}g^{\alpha\lambda}\Big(\frac{\partial
  g_{\lambda\mu}}{\partial x^{\nu}} + \frac{\partial
  g_{\lambda\nu}}{\partial x^{\mu}} - \frac{\partial
  g_{\mu\nu}}{\partial x^{\lambda}}\Big)
\end{eqnarray}
and ${\cal K}_{\sigma\mu\nu}$ is the contorsion tensor, related to
torsion ${\cal T}_{\sigma\mu\nu}$ by $ {\cal K}_{\sigma\mu\nu} =
\frac{1}{2}\,({\cal T}_{\sigma\mu\nu} + {\cal T}_{\mu\sigma\nu} +
     {\cal T}_{\nu\sigma\mu})$ and ${\cal T}^{\alpha}{}_{\mu\nu} =
     g^{\alpha \sigma}{\cal T}_{\sigma\mu\nu} =
     \Gamma^{\alpha}{}_{\mu\nu} - \Gamma^{\alpha}{}_{\nu\mu}$
     \cite{Schroedinger1950, Kostelecky2004}.  The integrand of the
     Einstein--Hilbert action Eq.(\ref{eq:1}) can be represented in
     the following form
\begin{eqnarray}\label{eq:5}
&&\sqrt{- g}\,{\cal R} = \sqrt{-g}\,R + \sqrt{-g}\,{\cal C} +
\frac{\partial}{\partial x^{\mu}}(\sqrt{-g}\,{{\cal K}^{\alpha}}_{\alpha\mu}) -
\sqrt{-g}\,g^{\mu\nu}\Big(\frac{1}{\sqrt{-g}}\,\frac{\partial}{\partial
  x^{\alpha}}(\sqrt{-g}\, {\cal K}^{\alpha}{}_{\nu\mu}) -
  \{^{\varphi}{}_{\alpha\mu}\}\,{\cal K}^{\alpha}{}_{\nu\varphi} -
  \{^{\alpha}{}_{\nu\varphi}\}\,{\cal K}^{\varphi}{}_{\alpha\mu}\Big),\nonumber\\ &&
\end{eqnarray}
where we have denoted 
\begin{eqnarray}\label{eq:6}
{\cal C} = g^{\mu\nu}({\cal K}^{\varphi}{}_{\alpha\mu}\,{\cal
  K}^{\alpha}{}_{\nu\varphi} - {\cal
  K}^{\alpha}{}_{\alpha\varphi}{\cal K}^{\varphi}{}_{\nu\mu})
\end{eqnarray}
and $R$ is the Ricci scalar curvature of the Einstein gravitational
theory, expressed in terms of the Christoffel symbols
$\{^{\alpha}{}_{\mu\nu}\}$ \cite{Rebhan2012} only.  Removing in
Eq.(\ref{eq:5}) the total derivatives and integrating by parts we
delete the third term and transcribe the fourth term into the form
$\sqrt{-g}\,{g^{\mu\nu}}_{;\alpha}\,{\cal K}^{\alpha}{}_{\nu\mu}$,
where ${g^{\mu\nu}}_{;\alpha}$ is the covariant derivative of the
metric tensor $g^{\mu\nu}$, vanishing because of the metricity
condition ${g^{\mu\nu}}_{;\alpha} = 0$ \cite{Rebhan2012}.

Since as has been shown in \cite{Ivanov2016a} that ${\cal C} = - 2
\Lambda_C$, where $\Lambda_C$ is the cosmological constant
\cite{Rebhan2012, Weinberg1989, Peebles1999b} (see also
\cite{Peebles2003}) or the relic dark energy density, the
Einstein--Hilbert action Eq.(\ref{eq:1}) of the Einstein--Cartan
gravitational theory with the scalar curvature Eq.(\ref{eq:2}) can be
represented in the following form \cite{Ivanov2016a}
\begin{eqnarray}\label{eq:7}
S_{\rm EH} = \frac{1}{2}\,M^2_{\rm Pl}\int d^4x\,\sqrt{-g}\,\big(R -
2\Lambda_C\big).
\end{eqnarray}
As has been shown in \cite{Ivanov2016a}, the same result is valid for
the Poincar\'e gauge gravitaitonal theory \cite{Kibble1961,
  Utiyama1956, Sciama1961,Blagojevic2001} (see also \cite{Hehl2012,
  Hehl2013, Hehl2019, Hehl2020}). Using Eq.(\ref{eq:7}) the action of
the Einstein--Cartan gravitational theory with torsion, chameleon
fields and matter fields we take in the form \cite{Ivanov2016a}
\begin{eqnarray}\label{eq:8}
S_{\rm EH} = \frac{1}{2}\,M^2_{\rm Pl}\,\int d^4x\,\sqrt{-g}\,R + \int
d^4x\,\sqrt{-g}\,{\cal L}[\phi] + \int d^4x\,\sqrt{- \tilde{g}}\,{\cal
  L}_m[\tilde{g}],
\end{eqnarray}
where ${\cal L}[\phi]$ is the Lagrangian of the chameleon field
\begin{eqnarray}\label{eq:9}
{\cal L}[\phi] =
\frac{1}{2}\,g^{\mu\nu}\,\partial_{\mu}\phi\partial_{\nu}\phi -
V(\phi)
\end{eqnarray}
and $V(\phi)$ is the potential of the chameleon self--interaction. In
Eq.(\ref{eq:8}), following Khoury and Weltman \cite{Khoury2004}, we
have included additively the cosmological constant $\Lambda_C$ in the
form of the relic dark energy density $\rho_{\Lambda} = M^2_{\rm
  Pl}\Lambda_C$ into the potential $V(\phi)$ of the chameleon field
self--interaction, i.e. $V(\phi) = \rho_{\Lambda} + \Phi(\phi)$.  This
implies that the chameleon field has no relation to the origin of the
cosmological constant or the relic dark energy density. It can only
evolve above the relic background of the dark energy, caused by
torsion.

The matter fields as well as the radiation \cite{Davis2009, Baum2014}
are described by the Lagrangian ${\cal L}_m[\tilde{g}_{\mu\nu}]$. The
interaction of the matter fields and radiation with the chameleon
field are expressed in terms of the metric tensor $\tilde{g}_{\mu\nu}$
in the Jordan frame \cite{Khoury2004,Mota2007,Dicke1962}, which is
conformally related to the Einstein--frame metric tensor $g_{\mu\nu}$
by $\tilde{g}_{\mu\nu} = f^2\,g_{\mu\nu}$ (or $\tilde{g}^{\mu\nu} =
f^{-2}\,g^{\mu\nu}$) and $\sqrt{- \tilde{g}} = f^4 \,\sqrt{-g}$ with
$f = e^{\,\beta\phi/M_{\rm Pl}}$, where $\beta$ is the
chameleon--matter coupling constant \cite{Khoury2004,Mota2007}. The
factor $f = e^{\,\beta\phi/M_{\rm Pl}}$ can be interpreted also as a
conformal coupling to matter fields and radiation \cite{Dicke1962}
(see also \cite{Khoury2004,Mota2007} and \cite{Ivanov2015}). For the
simplicity we have set the chameleon--photon coupling constant
$\beta_{\gamma}$ \cite{Baum2014} equal to the chameleon--matter
coupling constant $\beta$.

Varying the action Eq.(\ref{eq:8}) with respect to the metric tensor
$\delta g^{\mu\nu}$ (see, for example, \cite{Rebhan2012}) we arrive at
the Einstein equations, modified by the contribution of the chameleon
field. We get
\begin{eqnarray}\label{eq:10}
R_{\mu\nu} - \frac{1}{2}\,g_{\mu\nu}\,R =  -
\frac{1}{M^2_{\rm Pl}}\Big(f^2\,\tilde{T}^{(m)}_{\mu\nu} +
T^{(\phi)}_{\mu\nu}\Big),
\end{eqnarray}
where $R_{\mu\nu}$ is the Ricci tensor \cite{Rebhan2012},
$\tilde{T}^{(m)}_{\mu\nu}$ and $T^{(\phi)}_{\mu\nu}$ are the matter
(with radiation \cite{Rad2020, Tomas1930, Weinberg1971, Straumann1976,
  Schweizer1982, Schweizer1988}) and chameleon energy--momentum
tensors, respectively, determined by
\begin{eqnarray}\label{eq:11}
\tilde{T}^{(m)}_{\mu\nu} &=& \frac{2}{\sqrt{-
    \tilde{g}}}\,\frac{\delta }{\delta
  \tilde{g}^{\mu\nu}}\Big(\sqrt{-\tilde{g}}\,{\cal L}[\tilde{g}]\Big)
= (\tilde{\rho} + \tilde{p})\,\tilde{u}_{\mu}\tilde{u}_{\nu} -
\tilde{p}\,\tilde{g}_{\mu\nu},\nonumber\\ T^{(\phi)}_{\mu\nu} &=&
\frac{2}{\sqrt{- g}}\,\frac{\delta }{\delta g^{\mu\nu}}\Big(\sqrt{-
  g}\,{\cal L}[\phi]\Big) = \frac{\partial \phi}{\partial
  x^{\mu}}\,\frac{\partial \phi}{\partial x^{\nu}} -
g_{\mu\nu}\,\Big(\frac{1}{2}\,g^{\lambda\rho}\,\frac{\partial
  \phi}{\partial x^{\lambda}}\frac{\partial \phi}{\partial x^{\rho}} -
V(\phi)\Big).
\end{eqnarray}
The factor $f^2$ appears in front of $\tilde{T}^{(m)}_{\mu\nu}$
because of the relation
\begin{eqnarray}\label{eq:12}
\frac{2}{\sqrt{- g}}\,\frac{\delta }{\delta g^{\mu\nu}}\Big(\sqrt{-
  \tilde{g}}\,{\cal L}_m[\tilde{g}]\Big) = \frac{\sqrt{-
    \tilde{g}}}{\sqrt{- g}}\,\frac{\delta
  \tilde{g}^{\lambda\rho}}{\delta
  g^{\mu\nu}}\,\tilde{T}^{(m)}_{\lambda\rho} =
f^2\,\tilde{T}^{(m)}_{\mu\nu},
\end{eqnarray}
where we have used that
\begin{eqnarray}\label{eq:13}
 \frac{\sqrt{- \tilde{g}}}{\sqrt{- g}} = f^4\quad,\quad \frac{\delta
   \tilde{g}^{\lambda\rho}}{\delta g^{\mu\nu}} = f^{-2}
 \frac{1}{2}(g^{\lambda}{}_{\mu} g^{\rho}{}_{\nu} +
 g^{\lambda}{}_{\nu} g^{\rho}{}_{\mu}),
\end{eqnarray}
since $\tilde{g}^{\lambda\rho} = f^{-2}\,g^{\lambda\rho}$
\cite{Dicke1962} and $\tilde{T}^{(m)}_{\mu\nu} =
\tilde{T}^{(m)}_{\nu\mu}$. Then, the quantities $\tilde{\rho}$,
$\tilde{p}$ and $\tilde{u}_{\mu}$ in the Jordan frame are related to
the quantities $\rho$, $p$ and $u_{\mu}$ in the Einstein frame as
\cite{Dicke1962}
\begin{eqnarray}\label{eq:14}
\tilde{\rho} = f^{-3}\,\rho\quad,\quad\tilde{p} = f^{-3}\,p\quad,\quad
\tilde{u}_{\mu} = f\,u_{\mu}\quad,\quad\tilde{u}^{\mu} =
f^{-1}\,u^{\mu}.
\end{eqnarray}
This gives $\tilde{T}^{(m)}_{\mu\nu} = f^{-1} T^{(m)}_{\mu\nu}$.
Plugging Eqs.(\ref{eq:11}) with $\tilde{T}^{(m)}_{\mu\nu} = f^{-1}
T^{(m)}_{\mu\nu}$ into Eq.(\ref{eq:10}) we arrive at the Einstein
equations
\begin{eqnarray}\label{eq:15}
R_{\mu\nu} - \frac{1}{2}\,g_{\mu\nu}\,R = - \frac{1}{M^2_{\rm
    Pl}}\,T_{\mu\nu},
\end{eqnarray}
where $T_{\mu\nu}$ is the total energy--momentum tensor equal to
\begin{eqnarray}\label{eq:16}
T_{\mu\nu} = \Big((\rho + p)\,u_{\mu}u_{\nu} -
p\,g_{\mu\nu}\Big)\,e^{\,\beta\phi/M_{\rm Pl}} + \Big(\frac{\partial
  \phi}{\partial x^{\mu}} \frac{\partial \phi}{\partial x^{\nu}} -
g_{\mu\nu}\,\Big(g^{\lambda\rho}\,\frac{1}{2}\,\frac{\partial
  \phi}{\partial x^{\lambda}}\,\frac{\partial \phi}{\partial x^{\rho}}
- V(\phi)\Big)\Big),
\end{eqnarray}
where the contribution of torsion $T^{(\rm tor)}_{\mu\nu} =
\rho_{\Lambda} g_{\mu\nu} = M^2_{\rm Pl} \Lambda_C g_{\mu\nu}$
\cite{Ivanov2016a} is included additively to the potential $V(\phi)$
of the self--interactions of the chameleon field.  Below we analyze the
Einstein equations Eq.(\ref{eq:15}) in the Cold--Dark--Matter (CDM)
model \cite{PDG2020} in the Friedmann flat spacetime with the line
element \cite{Rebhan2012, PDG2020}
\begin{eqnarray}\label{eq:17}
ds^2 = g_{\mu\nu}(x) dx^{\mu}dx^{\nu} = dt^2 + a^2(t)\,\eta_{ij} dx^i dx^j ,
\end{eqnarray}
where $g_{00}(x) = 1$ and $g_{ij}(x) = a^2(t)\,\eta_{ij}$ with
$\eta_{ij} = - \delta_{ij}$. Then, $a(t)$ is the expansion parameter
of the Universe evolution \cite{Rebhan2012}. The Christoffel symbols
$\{^{\alpha}{}_{\mu\nu}\}$, the components of the Ricci tensor
$R_{\mu\nu}$ and the scalar curvature $R$ are equal to
\cite{Rebhan2012}
\begin{eqnarray}\label{eq:18}
\hspace{-0.3in} \{^0{}_{00}\} &=& \{^0{}_{0j}\} = \{^j{}_{00}\} =
\{^i{}_{kj}\} = 0,\;,\; \{^0{}_{kj}\} = - a\dot{a}\,\eta_{kj}\;,\;
\{^i{}_{0j}\} = \frac{\dot{a}}{a}\,\delta^i{}_j,\nonumber\\ R_{00} &=&
3\,\frac{\ddot{a}}{a}\;,\; R_{0j} = 0\;,\; R_{ij} =
\Big(\frac{\ddot{a}}{a} + 2 \frac{\dot{a}^2}{a^2}\Big)\,g_{ij}\;,\; R
= 6\,\Big(\frac{\ddot{a}}{a} + \frac{\dot{a}^2}{a^2}\Big),
\end{eqnarray}
where $\eta^{i\ell}\eta_{\ell j} = \delta^i{}_j$ and $\dot{a}$ and
$\ddot{a}$ are first and second derivatives with respect to time.

\section{Friedmann--Einstein equations of the Universe evolution}
\label{sec:evolution}

In the Friedmann spacetime the Einstein equations Eq.(\ref{eq:15})
define the equations of the Universe evolution, which are usually
called Friedmann's equations (or the Friedmann--Einstein equations)
\cite{Rebhan2012}. They are given by
\begin{eqnarray}\label{eq:19}
 \frac{\dot{a}^2}{a^2} = \frac{1}{3 M^2_{\rm Pl}}\,\big(\rho_{\phi} +
 (\rho_r + \rho_m)f(\phi) \big)
\end{eqnarray}
and 
\begin{eqnarray}\label{eq:20}
 \frac{\ddot{a}}{a} = - \frac{1}{6 M^2_{\rm Pl}}\,\big(\rho_{\phi} + 3
 p_{\phi} + (\rho_r + 3 p_r) f(\phi) + \rho_mf(\phi) \big),
\end{eqnarray}
where $\rho_r$ and $\rho_m$ are the radiation and matter
densities. The scalar field $\phi$ couples to radiation and matter
densities through the conformal factor $f(\phi) = e^{\,\beta\phi/M_{\rm
    Pl}}$. Then, the radiation density $\rho_r$ and pressure $p_r$ are
related by the equation of state $p_r = \rho_r/3$
\cite{Rebhan2012}. For the description of matter we use the
Cold Dark Matter (CDM) model with the pressureless
dark and baryon matter \cite{PDG2020}.  The scalar field density
$\rho_{\phi}$ and pressure $p_{\phi}$ are equal to
\begin{eqnarray}\label{eq:21}
\rho_{\phi} = \frac{1}{2}\,\dot{\phi}^2 + V(\phi)\quad,\quad p_{\phi}
= \frac{1}{2}\,\dot{\phi}^2 - V(\phi).
\end{eqnarray}
Varying the action Eq.(\ref{eq:8}) with respect to the scalar field
$\phi$ and its derivative one gets the equation of motion for the
scalar field \cite{Ivanov2015}. In the Friedmann spacetime it reads
\begin{eqnarray}\label{eq:22}
 \ddot{\phi} + 3\,\frac{\dot{a}}{a}\,\dot{\phi} + \frac{d
   V_{\rm eff}(\phi)}{d \phi} = 0,
\end{eqnarray}
where $V_{\rm eff}(\phi)$ is the effective potential given by
\begin{eqnarray}\label{eq:23}
 V_{\rm eff}(\phi) = V(\phi)  + \rho_m\big(f(\phi) - 1\big).
\end{eqnarray}
The contribution of the radiation density comes into the effective
potential in the form $(\rho_r - 3p_r)\big(f(\phi) - 1\big)$. Because
of the equation of state $p_r = \rho_r/3$ such a contribution
vanishes.  Thus, through the interaction with matter density $\rho_m$
the scalar field can acquire a non--vanishing mass if the effective
potential $V_{\rm eff}(\phi)$ obeys the constraints
\begin{eqnarray}\label{eq:24}
 \frac{dV_{\rm eff}(\phi)}{d\phi}\Big|_{\phi = \phi_{\rm min}} =
 0\quad,\quad \frac{d^2V_{\rm eff}(\phi)}{d\phi^2}\Big|_{\phi =
   \phi_{\rm min}} > 0, 
\end{eqnarray}
i.e. the effective potential $V_{\rm eff}(\phi)$ possesses a minimum
at $\phi = \phi_{\rm min}$. An important role for a dependence of a
chameleon field mass on a density of an environment plays the
conformal factor $f(\phi)$ and its deviation from unity.

\subsection{Bianchi identity, conservation of total energy--momentum 
tensor and conformal factor} 

Using Eq.(\ref{eq:18}) and taking into account that in
  the Friedmann flat spacetime the non--vanishing components of the
Einstein tensor $G^{\mu\nu} = R^{\mu\nu} - \frac{1}{2}\,g^{\mu\nu}R$
are equal to
\begin{eqnarray}\label{eq:25}
 G^{00} = - 3\,\frac{\dot{a}^2}{a^2}\quad,\quad G^{ij} = \Big(-
 2\,\frac{\ddot{a}}{a} - \frac{\dot{a}^2}{a^2}\Big)\,g^{ij}
\end{eqnarray}
one may show that the Einstein tensor $G_{\mu\nu}$ obeys the Bianchi
identity \cite{Rebhan2012}
\begin{eqnarray}\label{eq:26}
 {G^{\mu\nu}}_{;\mu} = \frac{1}{\sqrt{- g}}\,\frac{\partial}{\partial
   x^{\rho}}\Big(\sqrt{-g}\,G^{\rho\nu}\Big) +
 \Gamma^{\nu}{}_{\mu\rho} G^{\mu\rho} = 0,
\end{eqnarray}
where $G^{\mu\nu}{}_{;\mu}$ is a covariant divergence and
$\Gamma^{\nu}{}_{\mu\rho} = \{^{\nu}{}_{\mu\rho}\}$ are the
Christoffel symbols \cite{Rebhan2012}. As a result, the covariant
divergence of the total energy--momentum tensor $T^{\mu\nu}{}_{;\mu}$
should also vanish
\begin{eqnarray}\label{eq:27}
 {T^{\mu\nu}}_{;\mu} = \frac{1}{\sqrt{- g}}\,\frac{\partial}{\partial
   x^{\rho}}\Big(\sqrt{-g}\,T^{\rho\nu}\Big) +
 \Gamma^{\nu}{}_{\mu\rho}T^{\mu\rho} = 0.
\end{eqnarray}
Because of time--dependence only Eq.(\ref{eq:27}) takes the form
\begin{eqnarray}\label{eq:28}
 \frac{1}{\sqrt{- g}}\,\frac{\partial}{\partial
   t}\Big(\sqrt{-g}\,T^{00}\Big) +
      {\Gamma^0}_{ij}T^{ij} = 0,
\end{eqnarray} 
where we have taken into account Eq.(\ref{eq:18}). Using the
non--vanishing components of the total energy momentum tensor
\begin{eqnarray}\label{eq:29}
T^{00} = \rho_{\phi} + (\rho_r + \rho_m) f(\phi) \quad,\quad T^{ij} = -
\big(p_{\phi} + p_r f(\phi)\big)\,g^{ij}
\end{eqnarray}
we transcribe Eq.(\ref{eq:28}) into the form
\begin{eqnarray}\label{eq:30}
\frac{d}{dt}\Big(\rho_{\phi} + (\rho_r + \rho_m)\,f(\phi)\Big) +
3\,\frac{\dot{a}}{a}\Big(\rho_{\phi} + p_{\phi} + (\rho_r + p_r)\,f(\phi) +
\rho_m\,f(\phi)\Big) = 0.
\end{eqnarray}
Since Eq.(\ref{eq:22}) can be rewritten as follows 
\begin{eqnarray}\label{eq:31}
\frac{d}{dt}\Big(\rho_{\phi} + \rho_m\,f(\phi)\Big) = \frac{d}{dt}\rho_m -
3\,\frac{\dot{a}}{a}\,\Big(\rho_{\phi} + p_{\phi}\Big),
\end{eqnarray}
so we may remove the contribution of the chameleon field in
Eq.(\ref{eq:30}). As result, we get 
\begin{eqnarray}\label{eq:32}
\frac{d}{dt}\Big(\rho_rf(\phi) + \rho_m\Big) +
\frac{\dot{a}}{a}\,\Big(4 \rho_r f(\phi) + 3 \rho_m f(\phi)\Big) = 0,
\end{eqnarray}
where we have used the equation of state $p_r = \rho_r/3$
\cite{Rebhan2012}. Because of independence of radiation and matter
densities Eq.(\ref{eq:32}) can be splitted into evolution equations of
the radiation and matter densities
\begin{eqnarray}\label{eq:33}
\frac{d}{dt}\big(\rho_rf(\phi)\big) + 4 \frac{\dot{a}}{a}\big(\rho_r
f(\phi)\big) &=& 0,\nonumber\\ \frac{d}{dt}\rho_m + 3 \frac{\dot{a}}{a}\rho_m f(\phi)
&=& 0.
\end{eqnarray}
For the standard dependence of the radiation and matter densities on the
expansion parameter $a(t)$ \cite{Rebhan2012}
\begin{eqnarray}\label{eq:34}
 \rho_r = 3 M^2_{\rm Pl}{\rm H}^2_0
 \Omega_r\,\frac{a^4_0}{a^4}\quad,\quad \rho_m = 3 M^2_{\rm Pl}{\rm
   H}^2_0 \Omega_m\,\frac{a^3_0}{a^3},
\end{eqnarray}
where $a_0$, ${\rm H}_0 = 1.438(11) \times 10^{-33}\,{\rm eV}$,
$\Omega_r$ and $\Omega_m$ are the expansion parameter, the Hubble rate
and the relative radiation and matter densities at our time $t_0 =
1/{\rm H}_0$ \cite{PDG2020}, the equations for the radiation and
matter densities Eq.(\ref{eq:33}) are satisfied identically for
$f(\phi) = 1$.

Thus, if the radiation and matter densities depend on the expansion
parameter $a$ as $\rho_r(a) \sim a^{-4}$ and $\rho_m(a) \sim a^{-3}$,
local conservation of the total energy--momentum in the Universe can
be fulfilled if and only if the conformal factor $f(\phi)$, relating
the Einstein and Jordan frames and defining the chameleon--matter
coupling, is equal to unity, i.e. $f(\phi) = 1$. However, in this case
there is no influence of the chameleon field on the evolution of the
radiation and matter densities and a dependence of the chameleon field
mass on a density of its environment. In turn, for $f(\phi) \neq 1$
the evolution equations Eq.(\ref{eq:33}) admit some exact
solutions. It is convenient to search these solution in dependence of
the expansion parameter $a$. Treating the conformal factor $f(\phi)$
as a function of the expansion parameter $a$, i.e. setting $f(\phi) =
f(a) \neq 1$, the solutions to Eq.(\ref{eq:35}) can be given by
\begin{eqnarray}\label{eq:35}
 \rho_r(a) &=&
 \rho_{r0}\,\frac{a^4_0}{a^4}\,\frac{f(a_0)}{f(a)},\nonumber\\ \rho_m(a)
 &=& \rho_{m0}\,\frac{a^3_0}{a^3}\,\exp\Big(3\int^{a_0}_a \frac{f(a')
   - 1}{a'}\,da'\Big),
\end{eqnarray}
where $\rho_{r0} = 3 M^2_{\rm Pl}{\rm H}^2_0 \Omega_r$ and $\rho_{m0}
= 3 M^2_{\rm Pl}{\rm H}^2_0 \Omega_m$ are the radiation and matter
densities at out time $t_0 = 1/{\rm H}_0$ and $a(t_0) = a_0$, i.e. in
the era of the late--time acceleration of the Universe expansion or
the dark energy--dominated era. The integration constants of the first
order differential equations Eq.(\ref{eq:26}) are fixed by the
condition $\rho_r(a_0) = \rho_{r0}$ and $\rho_m(a_0) = \rho_{m0}$,
respectively \cite{Rebhan2012, PDG2020}. According to the solutions
Eq.(\ref{eq:35}), the chameleon field makes an influence on the
evolution of the radiation and matter densities.

As an example of the conformal factor we may use $f =
e^{\,\beta\varphi(a)/M_{\rm Pl}}$ \cite{Khoury2004,Mota2007}, where
$\varphi(a)$ is the chameleon field as a function of the expansion
parameter $a$ and the solution to Eq.(\ref{eq:22}), i.e. $\phi(t) =
\varphi(a)$. Keeping the linear order contributions in the
$\beta\varphi(a)/M_{\rm Pl}$ expansion we get
\begin{eqnarray}\label{eq:36}
 \rho_r(a) &=& \rho_{r0}\,\frac{a^4_0}{a^4}\,\Big(1 +
 \frac{\beta}{M_{\rm Pl}}(\varphi(a_0) -
 \varphi(a))\Big),\nonumber\\ \rho_m(a) &=&
 \rho_{m0}\frac{a^3_0}{a^3}\,\Big( 1 + 3\,\frac{\beta}{M_{\rm
     Pl}}\,\int^{a_0}_a \varphi(a')\,\frac{da'}{a'}\Big).
\end{eqnarray}
Thus, the deviations of the radiation and matter densities from their
standard behaviour $\rho_r(a) \sim a^{-4}$ and $\rho_m(a) \sim
a^{-3}$ are given by
\begin{eqnarray}\label{eq:37}
 \delta \rho_r(a) &=& \frac{\beta}{M_{\rm
     Pl}}\,\rho_{r0}\,\frac{a^4_0}{a^4}\,\Big(\varphi(a_0) -
 \varphi(a)\Big),\nonumber\\ \delta \rho_m(a) &=& 3
 \,\frac{\beta}{M_{\rm Pl}}\,\rho_{m0}\frac{a^3_0}{a^3}\int^{a_0}_a
 \varphi(a')\,\frac{da'}{a'}.
\end{eqnarray}
Some observations of deviations of the radiation and matter densities
in the Universe from their standard form might, in principle, evidence
an existence of the chameleon field. Nevertheless, we have to
emphasize that the contribution of the conformal factor to the
radiation and matter densities at our time is not practically
observable. It is seen from the solutions Eq.(\ref{eq:35}) that the
conformal factor affects the evolution of the radiation and matter
densities during the radiation-- and matter--dominated eras only.

\subsection{The Friedmann--Einstein equation Eq.(\ref{eq:19}) as the 
first integral of the Friedmann--Einstein equation Eq.(\ref{eq:20})}

It is well--known that without the chameleon field and for the
conformal factor $f(a) = 1$ the Friedmann--Einstein differential
equation for $\dot{a}^2/a^2$ is the first integral of the
Friedmann--Einstein differential equation for $\ddot{a}/a$
\cite{Rebhan2012}. However, such a property of Eq.(\ref{eq:19}) with
the chameleon field and the conformal factor $f(a) \neq 1$ to be the
first integral of Eq.(\ref{eq:20}) was not so far investigated and
proved in literature. In order to prove that Eq.(\ref{eq:19}) is the
first integral of Eq.(\ref{eq:20}) with the contribution of the
chameleon field and the conformal factor $f(a) \neq 1$ we rewrite
Eq.(\ref{eq:19}) as follows
\begin{eqnarray}\label{eq:38}
 \frac{\dot{a}^2}{a^2} = \frac{1}{3 M^2_{\rm Pl}}\,\big(\rho_{\rm ch} +
 \rho_rf + \rho_m\big),
\end{eqnarray}
where $\rho_{\rm ch} = \rho_{\phi} + \rho_m(f - 1) =
\frac{1}{2}\,\dot{\phi}^2 + V_{\rm eff}(\phi)$ is the chameleon field
density, given by Eq.(\ref{eq:21}) with the replacement $V(\phi) \to
V_{\rm eff}(\phi)$ (see Eq.(\ref{eq:23})).  In order to find
$\rho_{\rm ch}$ as a function of the expansion parameter $a$ we use
Eq.(\ref{eq:22}) and transcribe it into the form
\begin{eqnarray}\label{eq:39}
 a\frac{d}{da}\rho_{\rm ch}(a) + 6\rho_{\rm ch}(a) =  6 V_{\rm eff}(a),
\end{eqnarray}
where we have denoted $V_{\rm eff}(\phi) = V_{\rm eff}(a)$, assuming
that $\phi$ is a function of $a$, i.e. $\phi = \phi(a)$. As a
function of the expansion parameter $a$ the effective potential
$V_{\rm eff}(a)$ is given by 
\begin{eqnarray}\label{eq:40}
 V_{\rm eff}(a) = V(a) + \rho_m(a)(f(a) - 1),
\end{eqnarray}
where $V(a) = V(\phi) = V(\varphi(a))$ with the additive contribution
of the relic dark energy density, induced by torsion, and $f(a) =
e^{\,\beta\varphi(a)/M_{\rm Pl}}$. The solution to Eq.(\ref{eq:39}) is
equal to
\begin{eqnarray}\label{eq:41}
 \rho_{\rm ch}(a) = \frac{C_{\phi}}{a^6} + \frac{6}{a^6}\int a^5
 V_{\rm eff}(a)da,
\end{eqnarray}
where the term $C_{\phi}/a^6$ corresponds to the contribution of the
kinetic term of a scalar field \cite{Steinhardt2002}. The integration
constant $C_{\phi}$ we define as follows $C_{\phi} = 3M^2_{\rm Pl}{\rm
  H}^2_0 \Omega_{\phi}a^6_0$, where $\Omega_{\phi}$ is the integration
constant, having the meaning of a relative density of a scalar field
at time $t_0 = 1/{\rm H}_0$ \cite{PDG2020}. As a result,
Eq.(\ref{eq:38}) takes the form
\begin{eqnarray}\label{eq:42}
 \frac{\dot{a}^2}{a^2} = \frac{1}{3 M^2_{\rm Pl}}\,\big(\rho_{\rm ch}(a) +
 \rho_r(a)f(a) + \rho_m(a)\big),
\end{eqnarray}
where in the right-hand-side (r.h.s.) all densities and the conformal
factor are functions of the expansion parameter $a$.  Further it is
convenient to rewrite Eq.(\ref{eq:20}) as follows
\begin{eqnarray}\label{eq:43}
 \frac{\ddot{a}}{a} = - 2\frac{\dot{a}^2}{a^2} + \frac{1}{3M^2_{\rm
     Pl}}\,\rho_r(a) f(a) + \frac{1}{2M^2_{\rm Pl}}\,\rho_m(a) f(a) +
 \frac{1}{M^2_{\rm Pl}}\,V(a),
\end{eqnarray}
where we have used Eq.(\ref{eq:42}).  Since the second derivative
$\ddot{a}$ of the expansion parameter $a$ with respect to time can be
given by
\begin{eqnarray}\label{eq:44}
 \ddot{a} = \frac{1}{2}\,\frac{d\dot{a}^2}{da},
\end{eqnarray}
one may transcribe Eq.(\ref{eq:43}) into the form
\begin{eqnarray}\label{eq:45}
a\,\frac{d}{da}\dot{a}^2 + 4 \dot{a}^2 = \frac{2}{3M^2_{\rm
    Pl}}\,a^2\rho_r(a) f(a) + \frac{1}{M^2_{\rm Pl}}\,a^2 \rho_m(a)
f(a) + \frac{2}{M^2_{\rm Pl}}\,a^2 V(a).
\end{eqnarray}
The solution to Eq.(\ref{eq:45}) amounts to
\begin{eqnarray}\label{eq:46}
\dot{a}^2 = \frac{C}{a^4} + \frac{2}{3M^2_{\rm
    Pl}}\,\frac{1}{a^4}\int a^5\rho_r(a) f(a) da + \frac{1}{M^2_{\rm
    Pl}}\,\frac{1}{a^4}\int a^5 \rho_m(a) f(a)da + \frac{2}{M^2_{\rm
    Pl}}\,\frac{1}{a^4}\int a^5 V(a) da,
\end{eqnarray}
where $C$ is the integration constant. Dividing both sides of
Eq.(\ref{eq:46}) by $a^2$ we arrive at the equation
\begin{eqnarray}\label{eq:47}
\frac{\dot{a}^2}{a^2} = \frac{1}{3 M^2_{\rm
    Pl}}\Big(\frac{C_{\phi}}{a^6} + \frac{6}{a^6}\int a^5 U(a) da +
\frac{2}{a^6}\int a^5\rho_r(a) f(a) da + \frac{3}{a^6}\int a^5
\rho_m(a) f(a)da\Big),
\end{eqnarray}
where we have set $C_{\phi} = 3 M^2_{\rm Pl}C = 3 M^2_{\rm Pl} {\rm
  H}^2_0 \Omega_{\phi}$. Thus, Eq.(\ref{eq:47}) is the first integral
of Eq.(\ref{eq:20}). Making a replacement $V(a) = V_{\rm eff}(a) -
 \rho_m(a)(f(a) - 1)$ we arrive at the expression
\begin{eqnarray}\label{eq:48}
\frac{\dot{a}^2}{a^2} = \frac{1}{3 M^2_{\rm Pl}}\Big(\rho_{\rm ch}(a)
+ \frac{2}{a^6}\int a^5\rho_r(a)f(a) da + \frac{6}{a^6}\int
a^5\rho_m(a) da - \frac{3}{a^6}\int a^5 \rho_m(a) f(a)da\Big).
\end{eqnarray}
Since the radiation and matter densities as functions of $a$ obey the
equations
\begin{eqnarray}\label{eq:49}
a\frac{d}{da}\Big(\rho_r(a)f(a)\Big) &=& - 4 \Big(\rho_r(a) f(a)\Big),
\nonumber\\ a\frac{d}{da}\rho_m(a) &=& - 3 \rho_m(a) f(a)
\end{eqnarray}
and that $\rho_r(a)f(a) = \rho_{r0}f(a_0)a^4_0/a^4$ (see
Eq.(\ref{eq:35})), we transcribe the right--hand--side (r.h.s.) of
Eq.(\ref{eq:48}) into the form
\begin{eqnarray}\label{eq:50}
\frac{\dot{a}^2}{a^2} = \frac{1}{3 M^2_{\rm Pl}}\Big(\rho_{\rm ch}(a)
+ \rho_r(a)f(a) + \frac{1}{a^6}\int \frac{d}{da}\Big(a^6\rho_m(a)\Big)
da\Big) = \frac{1}{3 M^2_{\rm Pl}}\Big(\rho_{\rm ch}(a) +
\rho_r(a)f(a) + \rho_m(a)\Big),
\end{eqnarray}
This proves that Eq.(\ref{eq:19}) is the first integral of
Eq.(\ref{eq:20}) if the total energy--momentum is locally conserved.
The evolution of the chameleon field density $\rho_{\rm ch}(a)$ in
dependence of the expansion parameter $a$ is defined by
Eq.(\ref{eq:41}), which we rewrite as follows
\begin{eqnarray}\label{eq:51}
 \rho_{\rm ch}(a) = \rho_{\Lambda} + \frac{C_{\phi}}{a^6} +
 \frac{6}{a^6}\int a^5 \Phi(a)da + \frac{6}{a^6}\int a^5
 \rho_m(a)\big(f(a) - 1\big)\,da,
\end{eqnarray}
where $\rho_{\Lambda} = M^2_{\rm Pl} \Lambda_C$ and the third term in
Eq.(\ref{eq:51}) is the model--dependent part of the potential of the
self--interaction of the chameleon field $ V(\phi) = \rho_{\Lambda} +
\Phi(\phi)$ \cite{Ratra1988} -- \cite{Tsujikawa2013,
  Brax2004,Copeland2009}, taken as a function of the expansion
parameter $a$, i.e. $\Phi(\phi) = \Phi(a)$. Such a chameleon field
density may affect the acceleration of the Universe expansion. Setting
$f(a) = 1$ in Eq.(\ref{eq:51}) we get
\begin{eqnarray}\label{eq:52}
 \rho_{\rm ch}(a) = \rho_{\Lambda} + \frac{C_{\phi}}{a^6} +
 \frac{6}{a^6}\int a^5 \Phi(a)da, 
\end{eqnarray}
where the second and the last terms might still provide an
acceleration of the Universe expansion additional to that caused by
the first term $\rho_{\Lambda}$, which is induced by torsion
\cite{Ivanov2016a}. 

\section{Conclusion}
\label{sec:Schluss}

Having provided a geometrical origin for the cosmological constant or
the relic dark energy torsion had deprived the chameleon field to have
a chance to be quintessence or a hypothetical form of dark energy. As
a result the chameleon field is able only to evolve above the relic
background of the dark energy, caused by torsion, but not to originate
it. Then, as consequence of conservation of the total energy--momentum
of the system, the chameleon field can affect the dark energy dynamics
and as well as the Universe expansion even also the late--time
acceleration. We have shown that such an influence of the chameleon
field on the acceleration of the Universe expansion retains also even
if the conformal factor, relating the Einstein and Jordan frames and
defining the interaction of the chameleon field with its ambient
matter, is equal to unity (see Eq.(\ref{eq:51}) and
Eq.(\ref{eq:52}). This result is closely related to our proof that for
the system, including the chameleon field, radiation and matter (dark
and baryon matter), the Friedmann--Einstein equation for
$\dot{a}^2/a^2$ is the first integral for the Friedmann--Einstein
equation for $\ddot{a}/a$.

We have found that local conservation of the total energy--momentum of
the system, including the chameleon field, radiation and matter (dark
and baryon matter), leads to the equations of the evolution of the
radiation and matter densities, corrected by the conformal factor. Due
to the conformal factor, the radiative and matter densities as
functions of the expansion parameter $a$ differ from their standard
behaviour $\rho_r(a) \sim a^{-4}$ and $\rho_m(a) \sim a^{-3}$
\cite{Rebhan2012}. However, these deviations might be, in principle,
noticeable only during the radiation- and matter-dominated eras. In
the dark energy--dominated era that is in our time of the late-time
acceleration of the Universe, where the expansion parameter is equal
to $a_0 = a(t_0)$ for the Hubble time $t_0 = 1/H_0$ \cite{Rebhan2012},
the contributions of the conformal factor to the radiation and matter
densities in comparison to the standard values $\rho_r(a_0) = 3
M^2_{\rm Pl}{\rm H}^2_0 \Omega_r$ and $\rho_m(a_0) = 3 M^2_{\rm
  Pl}{\rm H}^2_0 \Omega_m$ are practically unobservable. This agrees
well with the constraints on the deviations of the radiation and
matter densities from their values at our time to a few parts per
million \cite{Jain2013a}, which can be obtained from the constraints
on the fifth force caused by the chameleon field in the Galaxy and the
Solar system.

The cosmological constant $\Lambda_C$, induced by torsion
\cite{Ivanov2016a}, we have included additively to the potential of
the self--interaction of the chameleon field as a background of the
relic dark energy: $V(\phi) = \rho_{\Lambda} + \Phi(\phi)$. In the
chameleon field theory \cite{Khoury2004, Mota2007} the relic dark
energy density $\rho_{\Lambda}$ is defined as follows $\rho_{\Lambda}
= \Lambda^4$, where the scale $\Lambda = \sqrt[4]{3 M^2_{\rm Pl}{\rm
    H}^2_0 \Omega_{\Lambda}} = 2.24(1)\,{\rm meV}$ is calculated for
the relative dark energy density $\Omega_{\Lambda} = 0.685(7)$
\cite{PDG2020}.  The $\phi$--dependent part of the potential of the
self--interaction of the chameleon field $\Phi(\phi)$ is arbitrary to
some extent, i.e. model--dependent, and demands a special analysis
similar to that carried out in \cite{Ratra1988} --
\cite{Tsujikawa2013, Brax2004, Copeland2009}. However, such an
analysis goes beyond the scope of our paper. We would like to
emphasize that a specific analysis of a dynamics of the chameleon
field such as different mechanisms of chameleon screening and a
formation of a fifth force, for example, in the Galaxy and the Solar
system is related also to a special choice of the potential of the
self--interaction of the chameleon field
\cite{Brax2004,Jain2013a}. Such an analysis has been carried out by
Brax {\it et al.} \cite{Brax2004} and Jain {\it et al.}
\cite{Jain2013a}. The repetition of such an analysis goes beyond the
scope of this paper.

As regards the assertion by Wang {\it et al.} \cite{Wang2012} and
Khoury \cite{Khoury2013} that the conformal factor is practically
constant during the Hubble time, one may argue the conformal factor
might be, in principle, practically constant (or better to say unity),
but such a behaviour of the conformal factor does not prohibit the
chameleon field, evolving above the relic dark energy background
induced by torsion, to take a certain part in dark energy dynamics
and, correspondingly, in the acceleration of the Universe expansion
(see Eq.(\ref{eq:51}) and Eq.(\ref{eq:52})) and even so in the
late--time acceleration of the Universe expansion.

\section{Acknowledgements}

We are grateful to Hartmut Abele for stimulating discussions and to
Philippe Brax and Alkistis Pourtsidou for fruitful discussions.  The
work of A. N. Ivanov was supported by the Austrian ``Fonds zur
F\"orderung der Wissenschaftlichen Forschung'' (FWF) under the
contracts P31702-N27 and P26636-N20, and ``Deutsche
F\"orderungsgemeinschaft'' (DFG) AB 128/5-2. The work of M. Wellenzohn
was supported by the MA 23 (FH-Call 16) under the project ``Photonik -
Stiftungsprofessur f\"ur Lehre''.

\end{document}